\documentclass[12pt]{article}
\usepackage{epsf}
\usepackage{epsfig}
\usepackage{amsfonts}
\usepackage{cite}

\def\tr{{\rm Tr}}
\newcommand{\bea}{\begin{eqnarray}}
\newcommand{\eea}{\end{eqnarray}}
\newcommand{\beq}{\begin{equation}}
\newcommand{\eeq}{\end{equation}}
\newcommand{\ba}{\begin{array}}
\newcommand{\ea}{\end{array}}

\def\nn{\nonumber}

\newcommand{\lsim}{\raisebox{-0.13cm}{~\shortstack{$<$ \\[-0.07cm] $\sim$}}~}
\newcommand{\gsim}{\raisebox{-0.13cm}{~\shortstack{$>$ \\[-0.07cm] $\sim$}}~}

\begin{document}
\def\yu{{\bf Y_u}}
\def\yde{{\bf Y_{d/e}}}
\def\yn{{\bf Y_{\nu}}}
\def\ye{{\bf Y_e}}
\def\yud{{\bf Y^\dagger_u}}
\def\yded{{\bf Y^\dagger_{d/e}}}
\def\ynd{{\bf Y^\dagger_{\nu}}}
\def\yed{{\bf Y^\dagger_e}}
\def\mps{{\bf m^2_{\tilde\Psi}}}
\def\mph{{\bf m^2_{\tilde\Phi}}}
\def\mnu{{\bf m^2_{\tilde N}}}
\def\mpst{{\bf m^{2{\sf T}}_{\tilde\Psi}}}
\def\mpht{{\bf m^{2{\sf T}}_{\tilde\Phi}}}
\def\mnut{{\bf m^{2{\sf T}}_{\tilde N}}}

\setcounter{page}{0}

\topmargin -1.0cm
\oddsidemargin -0.4cm
\evensidemargin -0.4cm
\pagestyle{empty}
\begin{flushright}
UG-FT-152/03\\
CAFPE-22/03\\
July 2003\\
rev. 18.11.03
\end{flushright}
\vspace*{5mm}
\begin{center}

{\Large\bf Quasi-Degenerate Neutrinos and Lepton Flavor}

{\Large\bf Violation in Supersymmetric Models}

\vspace{1.4cm}
{\sc J.I. Illana, M. Masip}\\
\vspace{.5cm}
{\it Centro Andaluz de 
F\'\i sica de Part\'\i culas Elementales (CAFPE) and}\\
{\it Departamento de F\'\i sica Te\'orica y del Cosmos}\\
{\it Universidad de Granada}\\
{\it E-18071 Granada, Spain}\\

\end{center}
\vspace{1.4cm}
\begin{abstract}

In supersymmetric (SUSY) models 
the misalignment between fermion and sfermion families
introduces unsuppressed flavor-changing processes.
Even if the mass parameters 
are chosen to give no flavor violation, family dependent
radiative corrections make this adjustment not stable.
We analyze the rate of $\ell\rightarrow\ell'\gamma$ in 
SUSY-GUT models with three quasi-degenerate neutrinos
and universal scalar masses at the Planck scale. 
We pay special attention to a recently proposed scenario 
where the low-energy neutrino mixings are generated 
from identical quark and lepton mixings at large 
scales. We show that: {\it (i)}~To take universal slepton 
masses at the GUT scale is a very poor approximation, 
even in {\it no-scale} models.
{\it (ii)} For large neutrino Yukawa couplings 
the decay $\mu\rightarrow e\gamma$ would be observed 
in the planned experiment at PSI.
{\it (iii)} For large values of $\tan\beta$ the tau coupling
gives important corrections, pushing 
$\mu\rightarrow e\gamma$ and $\tau\rightarrow \mu\gamma$
to accessible rates. In particular,
the non-observation of these processes in the near future
would exclude the scenario with unification of quark and
lepton mixing angles. 
{\it (iv)} The absence of lepton flavor violating decays in
upcoming experiments 
would imply a low value of $\tan\beta$,
small neutrino couplings, and large ($\gsim 250$ GeV) 
SUSY-breaking masses.

\end{abstract}

\centerline{PACS numbers: 12.60.Jv; 13.35.-r; 14.60.Pq; 14.60.St}

\vfill

\eject

\pagestyle{empty}
\setcounter{footnote}{0}
\pagestyle{plain}


\section{Introduction}

Supersymmetric (SUSY) extensions of the standard model 
introduce new sources of flavor non-conservation. 
Essentially, the three fermion families get SUSY masses 
from Yukawa interactions, whereas their scalar partners 
must get additional
SUSY-breaking masses with a different origin.
In general this will produce a misalignment between 
fermions and sfermions and then tree-level flavor-changing 
couplings to gauginos and higgsinos \cite{Gabbiani:1996hi}.

An acceptable SUSY-breaking mass matrix 
${\bf m^2}$ should
be (nearly) diagonal after the rotation diagonalizing the 
corresponding fermion 
matrix. The most economical solution is that 
${\bf m^2}\propto {\bf 1}$, and the
three scalar masses coincide at the tree level. In particular,
in the most popular scenario SUSY is broken in a hidden sector
only connected via gravitational 
interactions with the standard model. Being universal,
gravity could generate identical masses for the three sfermion
families near the Planck scale \cite{sugra}. Obviously, this universality
would be broken by radiative corrections 
\cite{misalignment,Polonsky:1994sr}, as the 
observed pattern of
fermion masses tells us that the interactions
of the three families with the Higgs fields are very different.

The renormalization-group (RG) corrections 
to universal SUSY-breaking squark masses and their implications 
on $K$ and $B$ physics have been extensively studied \cite{Goto:1998qv}.
In the quark sector the Yukawa couplings define a pattern 
with a heavy third family and small mixing angles with 
the lighter families. At low energies, in the basis of 
quark mass eigenstates the squark mass matrix is not diagonal,
with off-diagonal terms determined by this pattern. In particular,
the top-quark corrections on the light quark
sector are reduced by the small size of the mixings.\footnote{
See \cite{Ciuchini:2003rg} for a recent analysis of the 
correlation between quark and lepton 
flavor-changing processes in SUSY-GUTs. There it is shown 
that the strongest bounds on these models will be set by
upcoming experiments on lepton decays.}
Here we would like to study the leptonic sector.
Although charged-lepton and down-quark masses do not look 
too different, 
the results on neutrino oscillations (see \cite{Altarelli:2003vk} 
for a recent review) suggest a completely 
different pattern of Yukawa couplings. 
First of all, they have to accommodate large lepton mixings ($3\sigma$
limits from \cite{Gonzalez-Garcia:2003qf}):
\bea
0.47\ < &\sin\theta_{12}& <\ 0.67 \ , \nonumber\\
0.56\ < &\sin\theta_{23}& <\ 0.83 \ , \nonumber\\
        &\sin\theta_{13}& <\ 0.23 \ .
\label{angles}
\eea
Second, the observed mass differences \cite{Gonzalez-Garcia:2003qf}
\bea
&5.4\ <\ \Delta m^2_{21}/10^{-5}\mbox{ eV}^2\ <\ 10 
\quad\quad\mbox{or}\quad\quad
 14\ <\ \Delta m^2_{21}/10^{-5}\mbox{ eV}^2\ <\ 19 \ ,& \nonumber\\
&1.5\ <\ \Delta m^2_{32}/10^{-3}\mbox{ eV}^2\ <\ 3.9&
\label{masses}
\eea
can be realized in different ways: with hierarchical Yukawa 
couplings (mass differences of the order
of the larger mass), or with almost degenerate couplings (mass 
differences much smaller than masses). In addition, the couplings
can be large or small, since the 
values of the neutrino masses depend on the 
large masses $M$ of the 
{\it right-handed} neutrinos that appear in the 
{\it seesaw} mechanism \cite{seesaw}. In this way, a mass of 
order 1 eV can be generated by a coupling $Y_\nu\approx 1$ 
with $M\approx 10^{14}$ GeV or by $Y_\nu\approx 10^{-2}$
with $M\approx 10^{10}$ GeV. Another difference with the 
quark sector has to do with the presence of more complex phases. In the
neutrino sector all the low energy observables (three neutrino 
masses and three complex mixings) together with the masses of 
the three right-handed neutrinos do not fix completely the matrix
${\bf Y}_\nu$ of Yukawa couplings. There appears a complex orthogonal
matrix ${\bf R}$ \cite{Casas:2001sr} that is important in the 
physics above the large scale $M$ (for example, a complex ${\bf R}$ 
would be necessary to generate leptogenesis \cite{Branco:2002xf}).

In this paper we study the lepton flavor violating decays
$\ell\rightarrow\ell'\gamma$ in SUSY models with
universal scalar masses at the Planck scale. A first objective
is to evaluate the relevance of the corrections due to the
running of the masses between 
$M_P=M_{\rm Planck}/\sqrt{8\pi}\approx 2\times 10^{18}$~GeV 
and the GUT scale $M_X\approx 2\times10^{16}$. In particular,
we will compare the three branching ratios 
$\ell\rightarrow\ell'\gamma$
taking universal masses at $M_X$ and at $M_P$.
We will include the case where the scalar masses
vanish at $M_P$ and are generated by (flavor blind) gaugino
radiative corrections (the {\it no-scale} model). 
We will discuss three types of RG corrections introducing
lepton flavor violation:
corrections from the top quark Yukawa coupling, which 
affect the masses of the charged slepton singlets in SU(5) models; 
corrections from a large tau Yukawa coupling, which is the case
in models with large $\tan\beta$; and corrections from the 
neutrino Yukawa couplings, which dominate in the usual models
with universal masses at $M_X$. We show
that the second type of corrections (usually neglected in the
literature) can change the off-diagonal slepton masses in
a $\approx 10\%$ for universal masses at $M_X$ or by a factor
of order 2 for universal masses at $M_P$.

\begin{figure}
\centerline{\epsfig{file=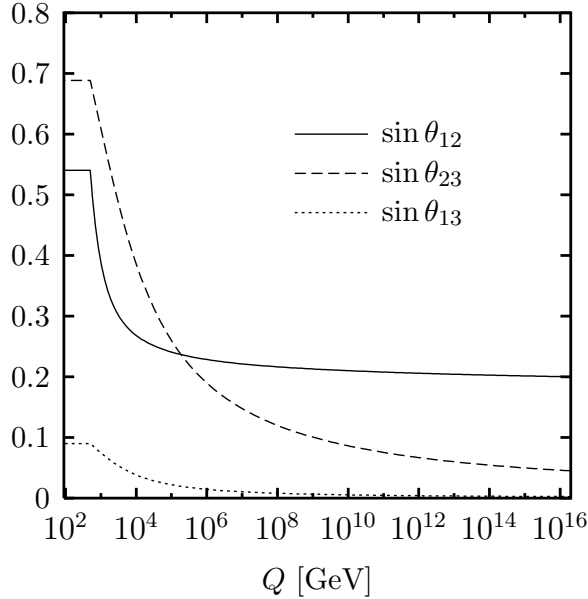,height=8cm}}
\caption{Evolution of lepton mixing angles from
CKM-like mixings at the GUT scale. The resulting 
low-energy angles are compatible with neutrino oscillation data 
($\tan\beta=50$).
\label{fig1}}
\end{figure}

A second objective in this paper is to analyze what is the
rate of lepton flavor violation in a particular model 
of neutrino masses and mixings recently proposed by 
Mohapatra {\it et al.} \cite{radmagnif}. They
observe that the 
angles required in these experiments can be
obtained from leptonic mixing angles identical to 
the ones in the quark sector at large scales (see Fig.~\ref{fig1}).
Although this possibility requires $\tan\beta\approx 50$ and
some amount of fine tunning 
(the degeneracy of the neutrino masses must increase with the
running, their values get {\it focused} in the infrared 
by tau corrections),
we think it provides a well motivated scenario. It suggests that
the dominant source of fermion mixing is in the interactions of 
down quarks and charged leptons, something 
natural in the simplest unification models (their couplings may
unify, for example, in SU(5)). The large values of the
tau Yukawa coupling required imply a violation
of flavor symmetry that could be in conflict with the data.

In our analysis we will consider quasi-degenerate neutrinos, with a
mass $m_\nu\approx 0.2$~eV that is compatible
with $m_\nu<2.2$~eV ($^3$H $\beta-$decay), $m_\nu<0.7/3$~eV (MAW) and
$m_\nu<(0.35-1.05)$~eV ($\beta\beta_{0\nu}$) \cite{absmass}.
We will assume that CP is conserved, with all the light neutrinos 
having the same CP parity and with a real matrix of Yukawa couplings. 
This implies a real matrix ${\bf R}$ and real 
neutrino mixings (it has been shown that a complex ${\bf R}$ may 
enhance the rate of lepton flavor violation in
several orders of magnitude \cite{Pascoli:2003rq}). 

For the mixing angles we will reproduce the
values in Eq.~(\ref{angles}), with $\theta_{13}$ below the 
bounds provided by the CHOOZ 
\cite{Apollonio:2002gd} and Palo Verde \cite{Boehm:2001ik} experiments. 
One should note, however, that very small 
values of this angle are not stable under radiative corrections
\cite{Casas:1999ac}.
Taking a zero value at $M_X$ we obtain at low energies values
that go from $\theta_{13}\approx 0.1$ for large $\tan\beta$
to $\theta_{13}\approx 5\times10^{-5}$ for small $\tan\beta$ and 
low $M$ ({\it i.e.} small
values of all the lepton Yukawa couplings).

\section{Radiative corrections to slepton masses}

Universal SUSY-breaking masses generated by gravitational
interactions receive radiative  
corrections from Yukawa interactions. In the minimal supersymmetric
standard model (MSSM) with right-handed neutrinos the relevant 
trilinear couplings in the superpotential are\footnote{
The contraction of SU(2) indices of two
doublets is $AB\equiv\epsilon_{ab} A^aB^b$, where $\epsilon_{ab}$
is the antisymmetric tensor.}
\beq
{\cal W}={Y_e}^{ij}\ E^c_{i} H_1  L_j
        +{Y_\nu}^{ij}\ N^c_{i} H_2 L_j\ ,
\label{2.1}
\eeq
where $L_i$, $E^c_i$ and $N^c_i$ stand for the three families of
lepton doublets, charged singlets, and neutrino singlets, respectively. 
$H_1$ and $H_2$ are the MSSM Higgs-doublet 
superfields. We will assume that the
neutrino singlets get a common mass at an intermediate scale $M$. 
Notice that the 
strength of the couplings necessary to reproduce the light neutrino
mass spectrum depends on $M$, becoming
top-quark-like when $M\approx 10^{14}$ GeV.\footnote{
The seesaw implies $Y_\nu\sim\sqrt{m_\nu M}/(v\sin\beta)$.}

{\it (i) Universal scalar masses at $M_X$}

Most previous analyses of the RG corrections 
\cite{Hisano:2001qz,Casas:2001sr,Carvalho:2001ex,Kageyama:2001zt,
Deppisch:2002vz} take universal slepton masses $ m_0$ at the GUT 
scale $M_X$, 
neglecting their running between 
$M_P$ and $M_X$ . We will compare the results
in that case with the ones in a minimal scenario
with SU(5) gauge symmetry between both scales, and will show
that the approximation (even in the no-scale case) is very poor.

We take at the GUT scale diagonal charged lepton couplings 
${\bf Y_e}$ and include all the lepton mixing in ${\bf Y_\nu}$.
Taking universal SUSY breaking masses\footnote{We neglect the 
scalar trilinears.} at $M_X$, the RG corrections introduce
two relevant effects (see {\it e.g.} 
\cite{Casas:2001sr,Castano:1993ri} for 
the RG equations of the MSSM with right-handed neutrinos). 
First, the running 
down to $M$ generates off-diagonal terms in the
slepton-doublet mass matrix:
\beq
m^2_{\tilde L\;ij}\approx -{3\over 8\pi^2}
m_0^2 ({\bf Y^{\dagger}_\nu Y_{\nu}})_{ij}\log{M_X \over M}\;.
\label{2.2}
\eeq
Second, the running also generates off-diagonal terms in 
${\bf Y_e}$. The low-energy charged-lepton mass matrix must be rediagonalized
with rotations $\theta^e_{12}$, $\theta^e_{13}$ and $\theta^e_{23}$ of order
\beq
\theta^e_{ij}\approx  -{1\over 16\pi^2}
({\bf Y^{\dagger}_\nu Y_{\nu}})_{ij}\log{M_X \over M}
\label{2.3}
\eeq
in the space of the three lepton doublets.
For large values of $\tan\beta$ the tau coupling separates
the third slepton family from the other two 
(it decreases $m^2_{{\tilde L}\;33}$ by a 20\%; see below).
When the rotation (\ref{2.3}) is performed also in the space of 
slepton doublets,
it induces new off-diagonal terms. More precisely, it gives a correction
(usually neglected in the literature) of order 
$1/(16\pi^2)Y_\tau^2 \log (M_X/M_Z)$ to the terms $m^2_{{\tilde L}\;i3}$
in Eq.~(\ref{2.2}).

To quantify the lepton-slepton misalignment we show 
${\bf m^2_{\tilde L}}$ at the electroweak scale $M_Z$ for 
$m_0=300{\rm \;GeV}$ at $M_X$.
We set the common gaugino mass $m_{1/2}=300$ GeV and 
distinguish large and small values of 
$\tan\beta$ (50 and 3, respectively). We take in both cases large
values of ${\bf Y_\nu}$ (around 0.9 at $M_X$, corresponding to 
$M=10^{14}$ GeV) that reproduce the observed pattern
of neutrino mixings and mass differences with $m_i\approx 0.2$ eV, obtaining
\bea
{\bf m^2_{\tilde L}}&=&(353 {\rm\; GeV})^2 
\pmatrix{1&-10^{-4}&-2\times10^{-5}\cr
-10^{-4}&0.999&-5\times10^{-4}\cr
-2\times10^{-5}&-5\times10^{-4}&0.793\cr}\;\;\;\;\;\;(\tan\beta=50)\;,
\label{2.4}
\\
{\bf m^2_{\tilde L}}&=&(352 {\rm\; GeV})^2 
\pmatrix{1&-5\times10^{-5}&5\times10^{-5}\cr
-5\times10^{-5}&0.997&-3\times10^{-3}\cr
5\times10^{-5}&-3\times10^{-3}&0.996\cr}\;\;\;\;\;\;(\tan\beta=3)\;.
\label{2.5}
\eea

The differences in the off-diagonal terms are due to the different
mixings assumed at the GUT scale (CKM-like for $\tan\beta=50$ and 
bimaximal for $\tan\beta=3$), with a $10\%$ contribution to 
$m^2_{\tilde L\; i3}$ from the
rediagonalization of the charged lepton mass matrix in the case of
large $\tan\beta$. The differences in the diagonal terms are 
only due to tau corrections (negligible
for $\tan\beta=3$). 

{\it (ii) Universal scalar masses at $M_P$}

Now let us give an estimate of the running
that includes RG corrections between $M_P$ and $M_X$. We will
consider an SU(5) grand unification framework 
\cite{Polonsky:1994sr,Barbieri:1994pv,Baek:2001kh}, as it is the 
simplest possibility. 
The LFV effects that we will find are minimal in the
sense that the unified theory could contain other sizable 
family-dependent couplings
in addition to the ones required to embed the MSSM. We do not 
include threshold effects at the GUT scale \cite{Polonsky:1994sr}, 
which could be as well an important source of 
lepton flavor asymmetry \cite{Hisano:1998cx}.

In SUSY-SU(5) each generation of quark doublets, up-quark singlets 
and charged-lepton
singlets can be accommodated in the same ${\bf 10}$ irrep ($\Psi_i$) 
of the group, whereas lepton doublets and down-quark singlets 
would be in the ${\bf \overline 5}$ ($\Phi_i$).
We also need gauge singlets ($N^c_i$) to generate neutrino masses, 
and a vectorlike ${\bf 5}+{\bf \overline 5}$ ($H_2$ and $H_1$) 
to include the two standard Higgs 
doublets. Other vectorlike fermions or the Higgs representations
needed to break the GUT symmetry are not essential in our
calculation. Including just the three fermion families the
trilinear terms in the superpotential read\footnote{$\Psi 
\Psi H_2\equiv \epsilon_{abcde}\; \Psi^{ab} \Psi^{cd} H_2^{e}$,
$\Psi \Phi H_{1}\equiv \Psi^{ab} \Phi_{a} H_{1b}$ and 
$N^c \Phi H_2\equiv N^c \Phi_{a} H_2^{a}$, where $a,...,e=1,...,5$ are 
SU(5) indices and $\epsilon_{abcde}$ is
the totally antisymmetric tensor.} \cite{Hisano:1998fj}
\beq
{\cal W}_{\rm SU(5)} =
{1\over 4} Y_u^{ij}\; \Psi_i \Psi_j H_2 + 
\sqrt{2}\ Y_{d/e}^{ij}\; \Psi_i \Phi_{j} H_{1} + 
Y_{\nu}^{ij}\; N^c_i \Phi_{j} H_2 \;.
\eeq

At $M_X$ we do the matching of Yukawa couplings in the following
way. {\it (i)} First we diagonalize  ${\bf Y_d}$ and ${\bf Y_e}$, 
and include all the quark mixing 
in ${\bf Y_u}$ and all the lepton mixing in ${\bf Y_{\nu}}$.
{\it (ii)} The matrix ${\bf Y_u}$ (symmetrized through a rotation 
of the up quark singlets) is then matched to the analogous
matrix above $M_X$. {\it (iii)} We do {\it not}
assume tau-bottom unification. To define ${\bf Y_{d/e}}$ we 
determine the tau and the bottom masses at $M_X$ ($m_\tau^0$ and
$m_b^0$); if 
the tau mass is larger, as it is usually the case for both the
small and the large values of $\tan\beta$ that we consider,
we match ${\bf Y_{d/e}}$ to ${\bf Y_e}$.
The smaller
bottom mass would then be explained through mixing ($\theta$) 
of the bottom component in $\Phi_3$ with the bottom component in 
a $\Phi+\overline \Phi$, which would reduce its
coupling by a factor of $\cos\theta=m_b^0/m_\tau^0$. 
For intermediate values of $\tan\beta$ the bottom is heavier
than the tau  at $M_X$, so we would proceed in the opposite way.
The lighter charged-lepton 
and down-quark masses could be separated by higher dimensional 
operators, but the radiative corrections that they introduce are
irrelevant.

The matching of SUSY-breaking scalar masses is straightforward
for squark doublets, up squark singlets, charged slepton singlets 
(all of them equal to ${\bf m^2_{\tilde \Psi}}$ between $M_X$ and $M_P$) 
and sneutrino singlets (${\bf m^2_{\tilde N}}$).
For down squark singlets and slepton doublets we take into account
the mixing with vectorlike fields, which reduces the 
RG corrections of the Yukawa couplings on the bottom squark mass 
squared by a factor of $\cos^2\theta$. 

We give in the Appendix the RG equations between $M_X$ and $M_P$ 
of Yukawa couplings and SUSY-breaking masses. 
The running introduces three main effects on the flavor structure 
of the model, two of them add to the corrections described for
universal masses at $M_X$, whereas the third one is new.
At $M_X$ there appear new off-diagonal terms 
in the slepton-doublet mass matrix:
\beq
m^2_{\tilde \Phi\;ij}\approx -{3\over 8\pi^2}
m_0^2 ({\bf Y^{\dagger}_\nu Y_{\nu}})_{ij}\log{M_P \over M_X}\ .
\eeq
The second effect is a large mass separation of the third slepton
family produced by tau corrections: 
\beq
m^2_{\tilde \Phi\;33}-m^2_{\tilde \Phi\;ii}\approx -{3\over 2\pi^2}
m_0^2\  Y_\tau^2\ \log{M_P \over M_X}\ .
\eeq
This mass splitting will introduce off-diagonal mass terms once the 
charged-lepton Yukawa matrix is rediagonalized at
low energies.
The final (new) effect has to do with the slepton-singlet mass 
matrix. In the minimal SU(5) model this matrix coincides with the
one for up squarks, so it will be affected by large top quark 
radiative corrections. At $M_X$ there will be 
off-diagonal terms of order
\beq
m^2_{\tilde \Psi\;ij}\approx -{9\over 8\pi^2}
m_0^2 ({\bf Y^{\dagger}_u Y_u})_{ij}\log{M_P \over M_X}\ .
\eeq
where ${\bf Y_u}$ contains the whole CKM rotation (we take
${\bf Y_{d/e}}$ diagonal at $M_X$). This effect was first considered
in \cite{Barbieri:1994pv}.

To illustrate the relevance of the corrections between $M_P$ and
$M_X$ we give
${\bf m^2_{\tilde L}}$ at $M_Z$ and compare it
with the matrices obtained for universal scalar masses 
at $M_X$. We take $m_{1/2}=275$ GeV and $m_0=300$ GeV 
at $M_P$, which give at $M_X$ similar values to 
the ones used in Eqs.~(\ref{2.4},\ref{2.5}):
\bea
{\bf m^2_{\tilde L}}&=&(353 {\rm\; GeV})^2 
\pmatrix{1&-4\times10^{-4}&-7\times10^{-5}\cr
-4\times10^{-4}&0.997&-2\times10^{-3}\cr
-7\times10^{-5}&-2\times10^{-3}&0.567\cr}\;\;\;\;\;\;(\tan\beta=50)\;,
\\
{\bf m^2_{\tilde L}}&=&(349 {\rm\; GeV})^2 
\pmatrix{1&-2\times10^{-4}&2\times10^{-4}\cr
-2\times10^{-4}&0.990&-10^{-2}\cr
2\times10^{-4}&-10^{-2}&0.989\cr}\;\;\;\;\;\;(\tan\beta=3)\;.
\eea

\begin{figure}
\begin{center}
\begin{tabular}{cc}
\epsfig{file=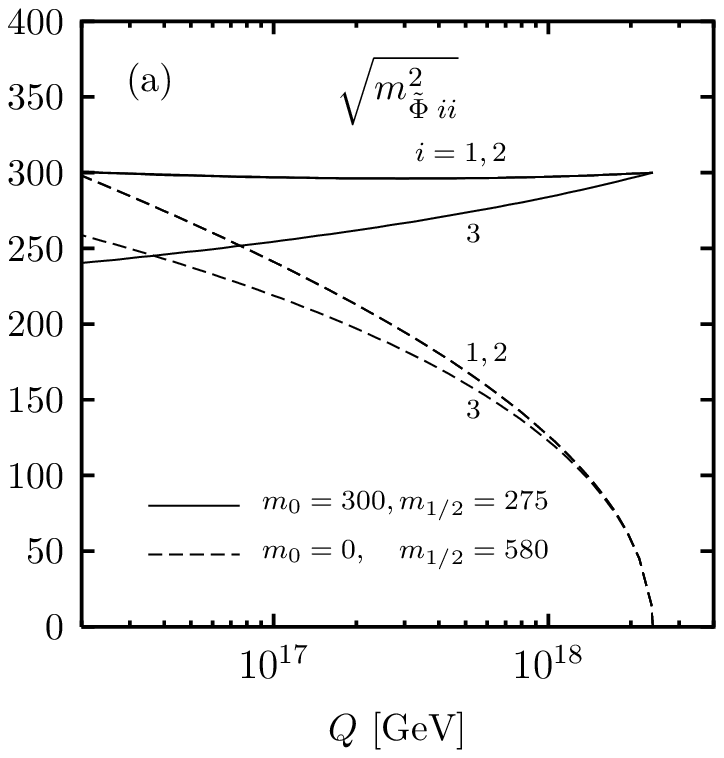,height=8cm} &
\epsfig{file=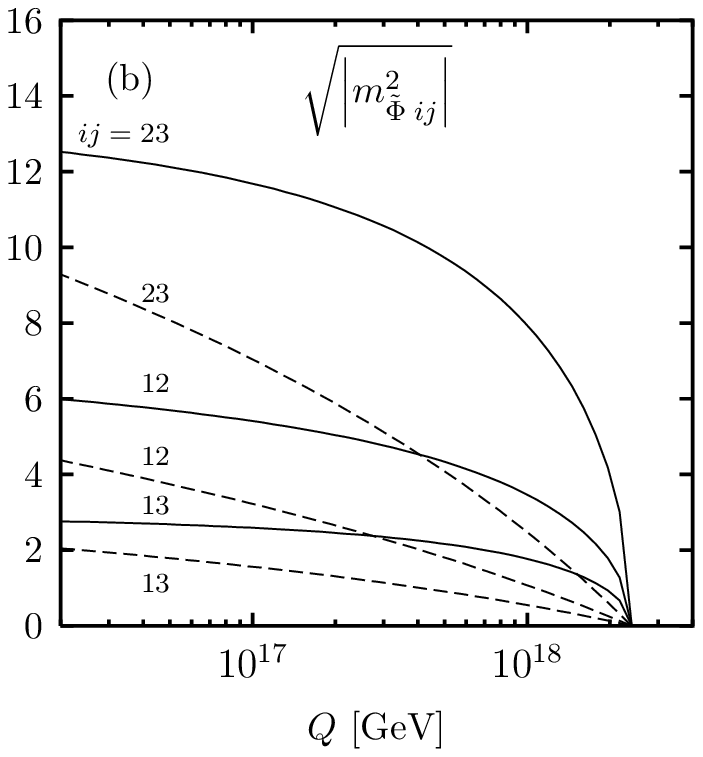,height=8cm} 
\end{tabular}
\end{center}
\caption{Mass splittings and off-diagonal terms [GeV] generated in 
the running of $\mph$ 
from $M_P$ to $M_X$. The diagonal terms for the first
and second families coincide. Dashed lines correspond to 
no-scale models ($m_0=0$ at $M_P$).
\label{fig2}}
\end{figure}

A final comment concerns the no-scale models. In
these models all scalar SUSY-breaking masses are zero at the Planck 
scale and non-zero values are generated mainly by 
{\it flavor-blind} gaugino 
corrections (see the RG equations in the Appendix). Once
the scalar masses are generated, however, family-dependent 
Yukawas will separate them.
Therefore, it is not clear whether or not it is justified
to take universal masses at $M_X$ in this scenario. Contrary to
the usual claim, we find that the mass splittings and the 
off-diagonal 
terms in the scalar sector 
at $M_X$ are only reduced by a factor of 
$\approx 0.8$ if the initial masses are taken zero at $M_P$.
We show in Fig.\ref{fig2}a 
$\sqrt{m^2_{\tilde \Phi\;ii}}$ at $M_X$ 
for $m_0=300$ GeV and $m_{1/2}=275$ GeV and and for 
$m_0=0$ GeV and $m_{1/2}=580$ GeV at $M_P$, taking $M=10^{14}$~GeV 
and a large value of $\tan\beta$. In the first case we obtain 
$\sqrt{m^2_{\tilde\Phi\;11}}\approx \sqrt{m^2_{\tilde\Phi\;22}} = 300$ 
GeV, and $\sqrt{m^2_{\tilde\Phi\;33}} = 240$ GeV, whereas in the no-scale
case we get 
$\sqrt{m^2_{\tilde\Phi\;11}}\approx \sqrt{m^2_{\tilde\Phi\;22}} = 300$ 
GeV, and $\sqrt{m^2_{\tilde\Phi\;33}} = 260$ GeV. 
In Fig.\ref{fig2}b we plot $\sqrt{|m^2_{\tilde\Psi\;i3}|}$ in the same
two cases, with qualitatively the same behaviour for this parameter.
We conclude that the
no-scale possibility does {\it not} justify taking universal 
SUSY-breaking scalar masses at the GUT scale $M_X$.

\section{Lepton flavor violation: 
$\mu\rightarrow e\gamma$, $\tau\rightarrow e\gamma$, 
$\tau\rightarrow \mu\gamma$}

Let us now determine how the misalignment between lepton and
slepton families translates into flavor-violating decays. The 
present experimental limits are
\bea
{\rm BR}(\mu\to e\gamma)   & < & 
1.2\times 10^{-11} \quad\mbox{\cite{Brooks:1999pu}} \ ,
\\
{\rm BR}(\tau\to e\gamma)  & < & 
2.7\times 10^{-6} \quad\mbox{\cite{Edwards:1996te}} \ ,
\\
{\rm BR}(\tau\to\mu\gamma) & < & 
6\times 10^{-7}   \quad\mbox{\cite{Inami:2002us}} \ ,
\eea
whereas future searches will be sensitive to branching ratios
of up to
\bea
{\rm BR}(\mu\to e\gamma)   & < & 10^{-14} \quad\mbox{\cite{meg}}\ ,
\\
{\rm BR}(\tau\to\mu\gamma) & < & 10^{-9}  \quad\mbox{\cite{Oshima:2001sh}}\ .
\eea

A detailed description of all the diagrammatics involved in these
processes can be found in \cite{Illana:2002tg}. Here we will just apply that 
calculation to the particular set of parameters described in the
previous section. 

We will distinguish four cases: large ({\it a}) and small ({\it b}) 
values of $\tan\beta$ (50 and 3, respectively), and large ({\it 1}) 
and small ({\it 2}) values of the neutrino Yukawa 
couplings (corresponding to values
of $M$ of $10^{14}$ and $5\times10^{11}$ GeV, respectively).
As explained in the introduction, the large values of $\tan\beta$
define a consistent scenario with unification of 
the quark and lepton mixing angles at $M_X$. In the cases with 
$\tan\beta=3$ we take a bimaximal mixing with $\theta_{13}=0$ at $M_X$,
which corresponds to low-energy values $\theta_{13}\approx 10^{-4}$.
Values of $\theta_{13}$ up to $10^{-2}$ give negligible effects
on the branching ratios, whereas larger values (up to the experimental
limit in Eq.~(\ref{angles})) give important corrections to the rates 
of $\mu\to e\gamma$ and $\tau\to e\gamma$.

{\it (i) Universal scalar masses at $M_X$}

In our numerical analysis we take a universal 
gaugino mass $m_{1/2}=100,300,500$ GeV 
at the GUT scale. 
In each case we vary the common slepton mass parameter 
$m_0$ at $M_X$ between 1~TeV and the
minimum value that gives acceptable masses for all
the slepton fields at low energies 
(we take the bounds from \cite{Hagiwara:fs}).
Notice that this minimum value strongly depends on the gaugino
mass parameter, as it gives important RG corrections that increase 
the low-energy value of the slepton masses.

\begin{figure}
\centerline{\epsfig{file=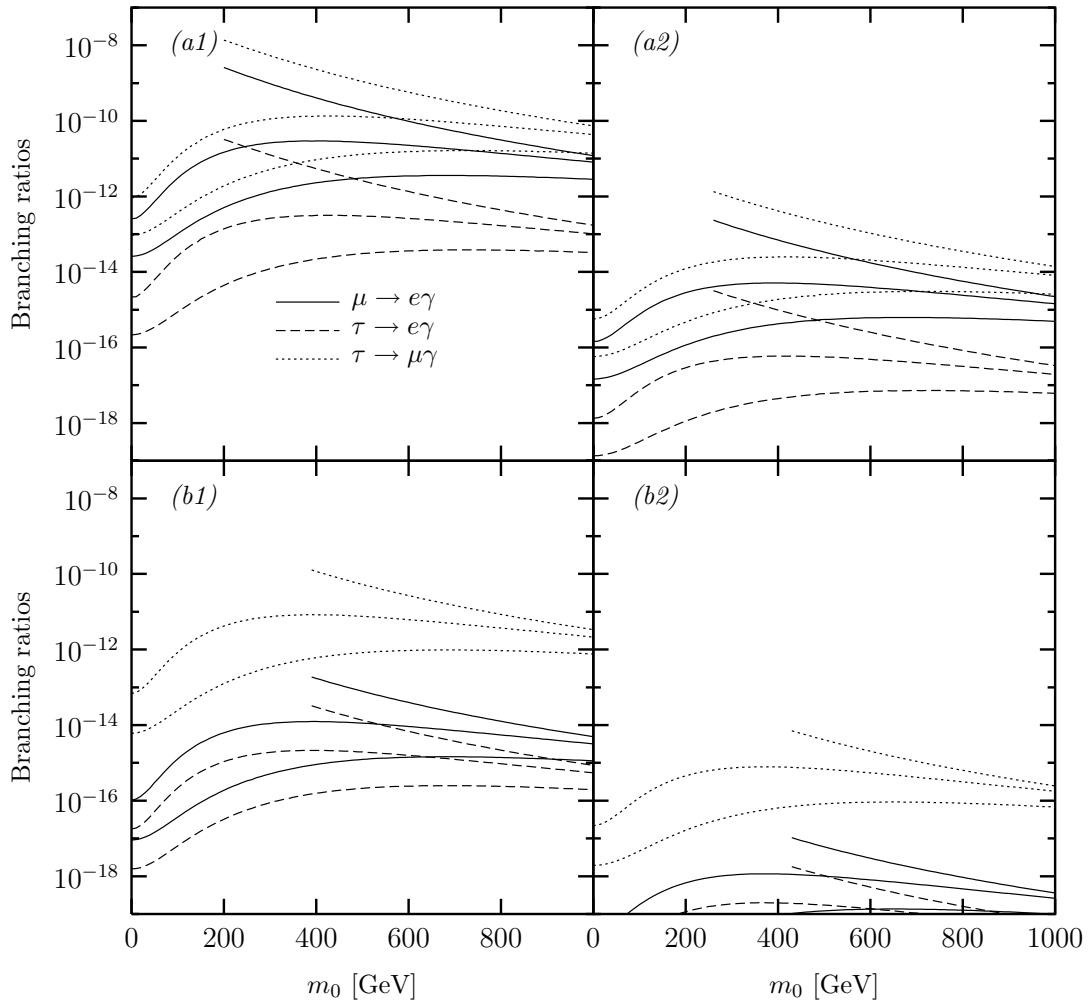,width=0.9\linewidth}}
\caption{
Branching ratios of $\ell\rightarrow\ell'\gamma$
for $m_{1/2}=100,300,500$ GeV and different values of the
scalar mass parameter $m_0$ at $M_X$. Cases {\it (a)}
and {\it (b)} correspond to $\tan\beta=50,3$, whereas
cases {\it (1)} and {\it (2)} correspond to 
$M=10^{14},5\times 10^{11}$ GeV, respectively. Lower values of 
$m_0$ for  $m_{1/2}=100$ GeV give slepton masses 
excluded by present bounds.
\label{fig3}}
\end{figure}

The results for the four cases  ({\it a1, a2, b1, b2})
are summarized in Fig.~\ref{fig3}. For each case we plot the branching 
ratios of $\mu\to e\gamma$ (solid), $\tau\to e\gamma$ (dashes) and 
$\tau\to \mu\gamma$ (dots). In all the cases 
the three branching ratios are dominated by diagrams with 
exchange of charginos and sneutrinos \cite{Illana:2002tg}.
The rates of $\mu\to e\gamma$ and $\tau\to e\gamma$ in the cases with low 
$\tan\beta$ would be scaled by a factor $\approx (\theta_{13}/10^{-2})^2$ 
for values of this angle close to the experimental bound 
$\theta_{13}\approx 0.2$.

{\it (ii) Universal scalar masses at $M_P$}

Now we take $m_{1/2}=100,300,500$ GeV at the Planck scale 
and universal scalar masses 
${\bf m^2_{\tilde \Phi}} = 
{\bf m^2_{\tilde \Psi}}=m^2_0 {\bf 1}$
also at the same scale. 
Again, we vary the parameter $m_0$ 
between 1~TeV and the minimum value not excluded experimentally.
The value $m_0=0$ at $M_P$, corresponding to no-scale models, 
is acceptable for large gaugino masses. 

\begin{figure}
\centerline{\epsfig{file=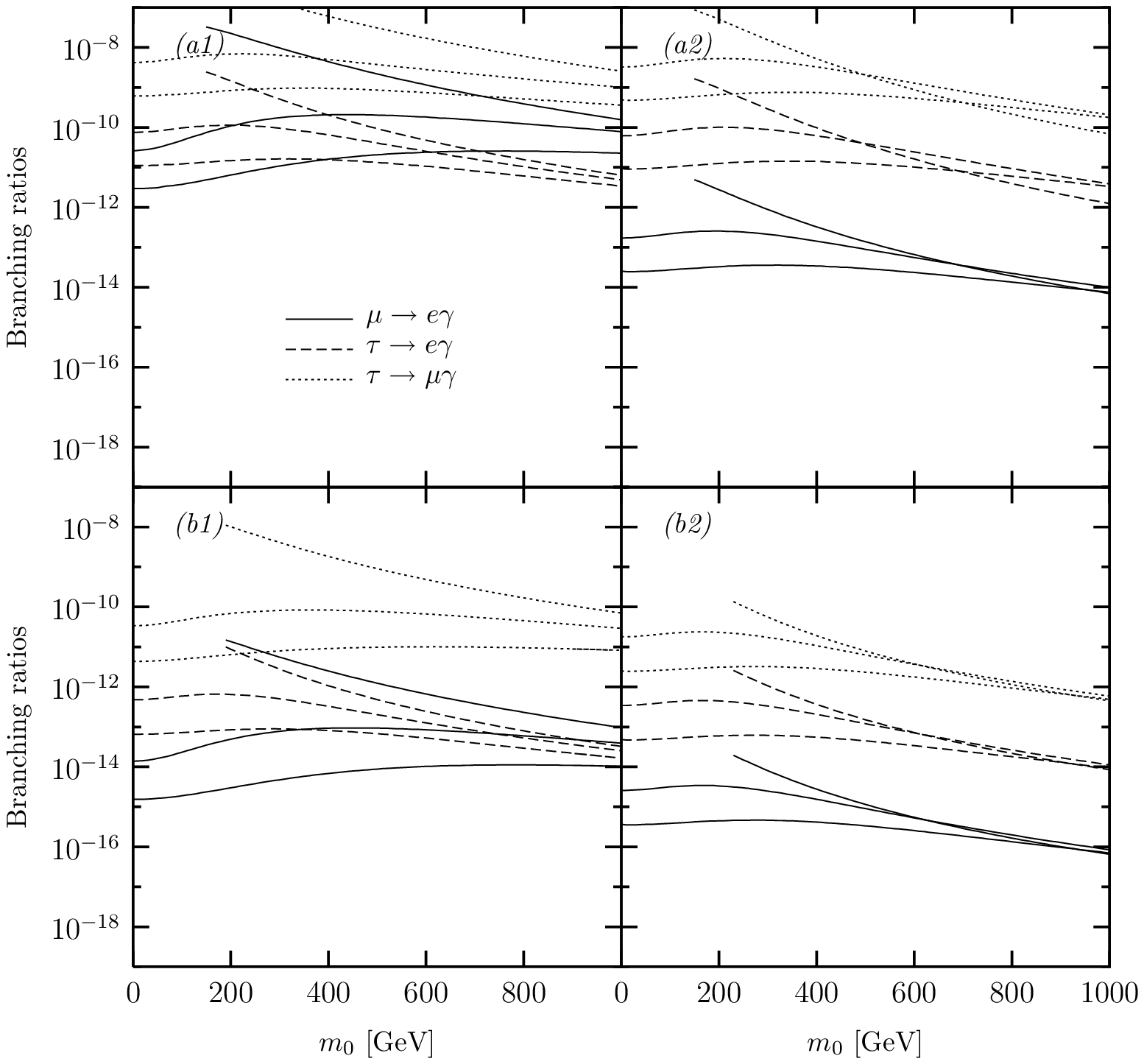,width=0.9\linewidth}}
\caption{Branching ratios of $\ell\rightarrow\ell'\gamma$
for $m_{1/2}=100,300,500$ GeV and different values of the
scalar mass parameter $m_0$ at $M_P$. Cases {\it (a)}
and {\it (b)} correspond to $\tan\beta=50,3$, whereas
cases {\it (1)} and {\it (2)} correspond to 
$M=10^{14},5\times 10^{11}$ GeV, respectively. $m_0=0$ defines
a no-scale scenario.
\label{fig4}}
\end{figure}

The results for the four cases described above are given in
Fig.~\ref{fig4}. The process $\mu\to e\gamma$ is dominated by 
chargino-sneutrino diagrams for large neutrino Yukawa
couplings (cases {\it a1} and {\it b1}) and by 
neutralino-slepton singlet diagrams otherwise. These second
diagrams also dominate in
$\tau\to \mu\gamma$ and $\tau\to e\gamma$ for all the values
of $\tan\beta$ and $M$ analyzed. The rate of $\mu\to e\gamma$
in the case {\it b1} would scale by a factor of 
$\approx (\theta_{13}/10^{-2})^2$
for larger values of this mixing angle, whereas the rest
of processes in case {\it b1} and the three processes in 
case {\it b2} would not change notably.

\section{Discussion}

The first question that we would like to address is 
how relevant are the RG corrections
between the Planck and the GUT scales. Comparing Figs.~\ref{fig3} and 
\ref{fig4} we find, for example, that these corrections can increase 
BR$(\tau\to \mu\gamma)$ in four orders of magnitude (case {\it a2})
or BR$(\mu\to e\gamma)$ in two orders of magnitude (case {\it b2}).
The approximation of equal slepton masses at $M_X$ is then 
clearly not justified. As discussed in Section 2, even in no-scale
models it gives a poor estimate of the slepton mass matrix
at low energies. 

We have shown that in the running between $M_P$ and $M_X$ 
both top quark and tau lepton corrections may be relevant. 
Top corrections affect the charged slepton singlet 
masses $m^2_{\tilde E^c\;ij}$, making the contributions 
mediated by slepton singlet and neutralino of
the same order as the ones mediated by slepton doublet and
chargino (which are proportional to $m^2_{\tilde L\;ij}$).
These corrections have been analyzed in the past 
\cite{Barbieri:1994pv,Hisano:1996qq} in some detail.
On the other hand, large tau corrections (in the 
large $\tan\beta$
regime) decrease the mass of the third slepton family, 
which introduces off-diagonal terms when the 
slepton matrix is rotated to the basis of charged-lepton 
mass eigenstates (see Eqs.~(5) and (10); the charged-lepton 
Yukawas get
off-diagonal terms in the running down to low energies). 
The analysis of tau corrections that we present here is 
absent in all previous
studies of  $\ell\rightarrow\ell'\gamma$, and it is relevant 
since it may amplify by a factor of 2 the rate 
of lepton flavor violation (notice that 
these corrections add to the linear scaling of the
amplitudes with $\tan^2\beta$ \cite{Hisano:1997tc}).

A second point in our work is to establish if the 
large $\tan\beta$ model proposed by 
Mohapatra {\it et al.} in \cite{radmagnif}
respects the present bounds on $\ell\rightarrow\ell'\gamma$,
and what would be the prospects at future experiments.
From cases {\it (a1) and (a2)} in Fig.~4 we see that
large values of the neutrino couplings ($M=10^{14}$ GeV)
would imply a BR$(\mu\to e\gamma)$ already excluded. On
the other hand, low values of the neutrino couplings
($M=5\times 10^{11}$ GeV)
would make the model consistent with all data, although
the non-observation of $\mu\to e\gamma$ 
at PSI \cite{meg} or of $\tau\to \mu\gamma$ 
at KEK \cite{Oshima:2001sh} 
would exclude this model.
These conclusions would be completely 
different if one neglects the 
running between $M_P$ and $M_X$ (see Fig.~3), as the model
could be consistent with the non observation of lepton
flavor violation at present and near future experiments.

From Fig.~4 we can extract as well how severe is the 
generic flavor problem of SUSY models in the lepton sector.
For low values of $\tan\beta$ the tau coupling is small, 
and we obtain (for $\theta_{13}<0.01$) 
BR$(\mu\to e\gamma)\lsim 10^{-11}$ 
(see the cases {\it b1} and {\it b2} in Fig.~\ref{fig4}),
a branching ratio that is within the present experimental limits.
Larger values of $\tan\beta$ will always put constraints on
the SUSY-breaking mass parameters. For example, the large values of
$\tan\beta$ required in \cite{radmagnif} respect the bounds on $\mu\to e\gamma$
only if the gaugino mass parameter $m_{1/2}$ is larger than 
500 GeV (case {\it a1} in Fig.~\ref{fig4}) or if the neutrino Yukawa
couplings are small enough, with $M\lsim 10^{11}$ GeV
(case {\it a2} in Fig.~\ref{fig4}).

A final point is whether these SUGRA models imply that 
lepton flavor-violating 
processes are necessarily
going to be observed in upcoming experiments. 
Again, both tau and neutrino Yukawa couplings play a relevant
role in the answer to that question. If all the lepton couplings
are small ({\it i.e.} for low values of $\tan\beta$ and $M$),
taking $m_{1/2}, m_0\gsim500$ GeV we find 
BR$(\mu\to e\gamma)\lsim 10^{-15}$, a value that could avoid
the $10^{-14}$ limit projected at PSI \cite{meg}. For lighter SUSY-breaking
masses or larger values of $\tan\beta$ or $M$
the gravity-mediated scenario 
of SUSY-breaking that we have considered 
predicts accessible rates of $\mu\to e\gamma$.
Another process with good experimental prospects is
$\tau\to \mu\gamma$. This process is sensitive to
the tau coupling and less dependent on neutrino couplings
(see the different cases in Fig.~\ref{fig4}). For $m_{1/2}=300$ GeV its
branching ratio goes from $10^{-8}$ for
large $\tan\beta$ to $10^{-11}$ for low values of 
$\tan\beta$ and $M$.
In summary, 
the non-observation of $\mu\to e\gamma$ and $\tau\to \mu\gamma$ 
would imply very low values of the Yukawas of the neutrinos 
{\it and} of $\tan\beta$. Comparing with the plots in
Fig.~3, 
we see that this generic conclusion can not be obtained if one assumes 
universality of the SUSY-breaking masses at $M_{GUT}$ 
instead of $M_{P}$.

\noindent{\bf Acknowledgements}\\
This work has been supported by the Spanish CICYT,
the Junta de Andaluc{\'\i}a and the European Union
under contracts FPA2000-1558, FQM 101,
and HPRN-CT-2000-00149, respectively.

\appendix
\section{RG corrections between $M_P$ and $M_X$}

The relevant RG equations between the Planck and the GUT  
scales are given below \cite{Baek:2001kh}. We neglect the scalar trilinears 
and define $t\equiv\log (Q/Q_0)$.

\begin{itemize}

\item Gauge coupling [$\alpha_G=g^2_5/(4\pi)$]:
\bea
\frac{d\alpha_G}{dt}=-\frac{3}{2\pi}\alpha^2_G\ .
\eea

\item Gaugino mass:
\bea
\frac{dM_5}{dt}=-\frac{3}{2\pi}\alpha_G M_5\ .
\eea

\item Yukawa couplings:
\bea
16\pi^2\frac{d}{dt}\yu&=&
	\yu
	\left[3\tr(\yud\yu)+\tr(\ynd\yn)
             +2\yded\yde+3\yud\yu\right]
\nn\\&+&
	\left(2\yded\yde+3\yud\yu\right)^{\sf T}\yu-\frac{96}{5}g^2_5\yu\ ,
\\
16\pi^2\frac{d}{dt}\yde&=&
	\yde
	\left[4\tr(\yded\yde)
             +2\yded\yde+3\yud\yu\right]
\nn\\&+&
	\left[2\yde\yded+3\left(\ynd\yn\right)^{\sf T}
		\right]\yde-\frac{84}{5}g^2_5\yde\ ,
\\
16\pi^2\frac{d}{dt}\yn&=&
	\yn
	\left[3\tr(\yud\yu)+\tr(\ynd\yn)
             +4\left(\yde\yded\right)^{\sf T}\right]
\nn\\&+&
             5\yn\ynd\yn-\frac{48}{5}g^2_5\yn\ .
\eea

\item SUSY-breaking masses:
\bea
8\pi^2\frac{d}{dt}\mps&=&
\mps\left(\yded\yde+\frac{3}{2}\yud\yu\right)
+\left(\yded\yde+\frac{3}{2}\yud\yu\right)\mps
\nn\\&+& 2\yded\left(\mpht+m^2_{H_1}
\right)\yde+3\yud\left(\mpst
+m^2_{H_2}\right)\yu
\nn-\frac{72}{5}g^2_5M^2_5\ ,
\\
8\pi^2\frac{d}{dt}\mph&=&
\mph\left[2(\yde\yded)^{\sf T}+\frac{1}{2}
\ynd\yn\right]+\left[2(\yde\yded)^{\sf T}+\frac{1}{2}
\yn\yn^\dagger\right]\mph\nn\\
&+& 4{\bf Y^*_{d/e}}(\mpst+m^2_{H_1})\yded+\ynd(\mnut
+m^2_{H_2})\yn-\frac{48}{5}g^2_5M^2_5\ ,
\\
8\pi^2\frac{d}{dt}\mnu&=&
\frac{5}{2}\mnu(\yn\ynd)^{\sf T}+\frac{5}{2}(\yn\ynd)^{\sf T}
\mnu+5{\bf Y^*_{\nu}}(\mpht+m^2_{H_2})
\ynd\ ,\quad
\\
8\pi^2\frac{d}{dt}m^2_{H_1}&=&
4m^2_{H_1}\tr(\yded\yde)+4\tr(\yde\mps\yded)
\nn\\&+&4\tr(\yded\mpht\yde)
-\frac{48}{5}M^2_5\ ,
\\
8\pi^2\frac{d}{dt}m^2_{H_2}&=&
m^2_{H_2}\left[3\tr(\yud\yu)+\tr(\ynd\yn)\right]
+6\tr(\yud\mpst\yu)
\nn\\&+&\tr(\ynd\mnut\yn)+\tr(\yn\mph\ynd)-\frac{48}{5}M^2_5\ .
\eea

\end{itemize}


\end{document}